\documentclass[aps,prd,twocolumn, preprintnumbers, floatfix, superscriptaddress,showpacs]{revtex4}
\usepackage[dvips]{graphics}
\usepackage{amsmath,amsfonts,bm}
\usepackage{epsfig,bm}
\usepackage{graphicx}
\begin{document}

\preprint{JLAB-THY-07-652}

\affiliation{Thomas Jefferson National Accelerator Facility,
             Newport News, VA 23606, USA}
\affiliation{Physics Department, Louisiana State University,
             Baton Rouge, LA 70803, USA}
\affiliation{Physics Department, Old Dominion University, Norfolk,
             VA 23529, USA}

\author{H.R. Grigoryan}
\affiliation{Thomas Jefferson National Accelerator Facility,
             Newport News, VA 23606, USA}
\affiliation{Physics Department, Louisiana State University,
             Baton Rouge, LA 70803, USA}
%\affiliation{Laboratory of Theoretical Physics, JINR, Dubna, Russian
%             Federation}
\author{A.V. Radyushkin}
\affiliation{Thomas Jefferson National Accelerator Facility,
              Newport News, VA 23606, USA}
\affiliation{Physics Department, Old Dominion University, Norfolk,
             VA 23529, USA}
\affiliation{Laboratory of Theoretical Physics, JINR, Dubna, Russian
             Federation}

\title{Structure of Vector Mesons in  Holographic Model with Linear Confinement}

\begin{abstract}
Wave functions and form factors of vector mesons  are  investigated in the holographic dual model of QCD with
oscillator-like infrared cutoff.  We   introduce wave functions conjugate to solutions of the 5D equation of
motion and develop a formalism based on these wave  functions,   which are very similar to those   of a
quantum-mechanical   oscillator. For the lowest bound state ($\rho$-meson),  we show that, in  this  model, the
basic elastic form factor  exhibits the  perfect vector meson dominance, i.e., it is    given by the $\rho$-pole
contribution  alone. The electric radius of the $\rho$-meson is   calculated, $\langle r_\rho^2 \rangle_C = 0.655
\, {\rm fm}^2$,  which is larger than in  case of the hard-wall cutoff. The  squared radii of higher excited
states   are found to increase logarithmically rather than linearly with the radial excitation number. We calculate the
coupling constant $f_\rho$ and find that the experimental value
is closer to that calculated in the  hard-wall model.
\end{abstract}

\maketitle

%======================================================================

\section{Introduction}

Holographic duals of quantum chromodynamics (QCD) which are based on the gauge/gravity correspondence
\cite{Maldacena:1997re} have been applied recently to hadronic physics  (see, e.g.,
\cite{Polchinski:2002jw,Boschi-Filho:2002vd,Brodsky:2003px,Sakai:2004cn,Erlich:2005qh,Erlich:2006hq,DaRold:2005zs,
Ghoroku:2005vt,Hirn:2005nr,Brodsky:2006uq,Karch:2006pv,Csaki:2006ji,Hambye:2005up}), and demonstrated their
ability to incorporate  such essential  properties of QCD as confinement and chiral symmetry breaking, and have
been successful   in many cases  in determination of static hadronic properties, i.e.,   resonance masses, decay
constants, chiral coefficients, etc. In Refs.~\cite{Polchinski:2002jw,Brodsky:2003px}, the dynamic properties
(form factors) have   been  studied within the holographic   approach of Ref.~\cite{Polchinski:2002jw}, and the
connection between AdS/QCD approach of Refs.~\cite{Polchinski:2002jw,Brodsky:2003px} and the usual light-cone
formalism  for  hadronic form factors was proposed in \cite{Brodsky:2006uq}  and  discussed in
\cite{Radyushkin:2006iz}. The calculation of form factors of scalar   and   vector hadrons within the approach of
Ref.~\cite{Polchinski:2002jw} was performed in Refs.~\cite{Hong:2004sa,Hong:2005np}, and   applied to   study
the universality  of   the $\rho$-meson couplings to   other   hadrons. In   our    recent paper
\cite{Grigoryan:2007vg}, we studied form factors and wave functions of vector   mesons   within the framework of
the holographic QCD model described in Refs.~\cite{Erlich:2006hq,Erlich:2005qh,DaRold:2005zs} (which will be
referred to as hard-wall model).

In the hard-wall model , the confinement is modeled by hard-wall cutting off the AdS space along the extra fifth dimension
at some finite value $z = z_0$. The solutions   of the relevant eigenvalue equation are   given by  the Bessel
functions,  and   masses of bound states are   given   by    the roots \mbox{$M_n =\gamma_{0,n}/z_0$} of
$J_0(Mz_0)$. As  a   result, the  masses of higher excitations behave like $ M^2_n \sim n^2 $. It  was argued
\cite{Karch:2006pv,Shifman:2005zn}
that, instead, one should expect $ M^2_n \sim n $ behavior. This connection can  be   derived
from semiclassical arguments \cite{Schreiber:2004ie,Shifman:2005zn}. An explicit  AdS/QCD  model which gives such
a  linear behavior was proposed in  Ref.~\cite{Karch:2006pv}. The hard-wall boundary conditions in  this model are
substituted by an oscillator-type potential providing a soft
IR cut-off in the action integral  (for  this  reason, it  will be  referred to as ``soft-wall model'').

The aim of the   present  paper is to   study form factors and wave functions of vector   mesons   within the
framework of   the soft-wall model  formulated in Ref.~\cite{Karch:2006pv}, and compare the results with those we obtained
in Ref.~\cite{Grigoryan:2007vg} investigating the hard-wall model . To this end, we extend the approach  developed in
Ref.~\cite{Grigoryan:2007vg}. We start with recalling, in Section II, the basics of the soft-wall model  and  some results
obtained in Ref.~\cite{Karch:2006pv}, in particular, the form of the relevant action, the eigenvalue equation for
bound   states  and its   solution.
In  Section III, we derive a useful integral representation for the  bulk-to-boundary propagator ${\cal V}(p,z)$
that allows to write  ${\cal V}(p,z)$ as an explicit  expansion over bound state poles with the $z$-dependence of
each pole contribution given by ``$\psi$ wave functions''  that are eigenfunctions of the 5D equation of motion.
Then we show that the same representation can be obtained from the general formalism of Green's functions.
However, as   we   already   emphasized in Ref.~\cite{Grigoryan:2007vg}, the $\psi_n(z)$ wave functions are not
direct analogues of the usual quantum-mechanical wave functions. In particular, a meson  coupling constant $f_n$
is obtained from the derivative of  $\psi_n(z)$ at $z=0$   rather   than   from   its value at  this point. To
this end,  we introduce  ``$\phi$ wave functions''  which look more like   wave functions of oscillator bound
states in  quantum mechanics.  Their values at $z=0$ give the  bound state couplings $g_5 f_n/M_n$,  they
exponentially decrease with $z^2$, and thus they have properties necessary  for the light-cone interpretation of
AdS/QCD results proposed in Ref.~\cite{Brodsky:2006uq}.
In Section IV, we study the three-point function $\langle JJJ \rangle$ and   obtain  expression for transition
form factors that involves  $\psi$ wave functions and the nonnormalizable mode factor ${\cal J}(Q,z)$.
The latter   is written  as a sum   over all bound states in
 the channel of electromagnetic current, which gives an analogue
of generalized vector meson dominance (VMD)  representation for hadronic  form factors.
In  Section V,  it is    shown   that it   is possible to rewrite form   factors in terms  of $\phi$ functions. Then we
formulate predictions for $\rho$-meson form factors,  and  analyze these predictions in the   regions of small and
large $Q^2$.
In particular, our formalism allows to calculate $\rho$-meson electric radius, and
 the radii of higher excited states.
It is also shown that, for the  basic $\rho$-meson form
factor ${\cal F} (Q^2)$  given   by   the   overlap of   the $\phi$ wave  functions, the soft-wall model  predicts
 exact
VMD pattern,  when  just one lowest bound state in the $Q^2$-channel  contributes. For another
 $\rho$-meson form factor
$F(Q^2)$, which is  given by the overlap of the $\psi$ wave  functions, a two-resonance dominance is   established, with  only
two  lowest bound states  in the $Q^2$-channel  contributing.
In Section VI, we compare our results  obtained in the soft-wall model  with  those
 derived in the hard-wall model  studies performed in Ref.~\cite{Grigoryan:2007vg}.
 At the end, we summarize the paper.

%======================================================================

\section{Preliminaries}

We consider the gravity background with a smooth cutoff that was proposed in
Ref.~\cite{Karch:2006pv} instead of a hard-wall infrared (IR) cutoff. In this case,
the only background fields are dilaton $ \chi(z) = z^2 \kappa^2$ and metric
$ g_{MN} $.
 The metric can be written as
\begin{align}
g_{MN}dx^Mdx^N = \frac{1}{z^2}\left(\eta_{\mu \nu}dx^{\mu}dx^{\nu} -
dz^2\right) \ ,
\end{align}
where $ \eta_{\mu\nu} = {\rm Diag}(1,-1,-1,-1) $ and $ \mu, \nu = (0,1,2,3) $, $ M, N = (0,1,2,3,z) $. To
determine the spectrum of vector mesons,  one needs the quadratic part of the action
\begin{align}
S_{\rm AdS} = - \frac{1}{4g_5^2}\int d^4x~\frac{dz}{z}\,
e^{-\chi} ~{\rm
Tr}\left(F_{MN}F^{MN}\right) \ ,
\end{align}
where $ F_{MN} = \partial_{M}V_{N} - \partial_{N}V_{M} - i
[V_{M},V_{N}] $, $ V_{M} = t^a V^a_M $, ($t^a= \sigma^a/2$,
with $\sigma^a$ being  Pauli matrices).
In the axial-like gauge $ V_z = 0$,  the  vector field   $ V^a_\mu (x, z=0)$ corresponds to the source for  the
vector current $ J^a_\mu(x) $. To obtain the equations of motion for the transverse component of the field,  it is
convenient to work with the Fourier transform $ \tilde V^a_\mu (p, z) $ of $ V^a_\mu (x, z)$,  for which one has
\begin{align}
\label{genbasiceq}
\left (\partial_z\left[\frac{1}{z}\, e^{-z^2}\partial_z \tilde V^a_\mu (p, z)\right] + p^2
\frac{1}{z}\, e^{-z^2}\tilde V^a_\mu (p, z) \right )_{\perp}  = 0 \   .
\end{align}
(Here, and in the rest of the paper, we find it convenient to follow the convention of Ref.~\cite{Karch:2006pv},
 in which the oscillator scale $\kappa$
is treated as 1, i.e., we write below $z^2$ instead of $\kappa^2 z^2$,
$e^{-z^2}$ instead of $e^{-z^2\kappa^2}$, etc.  Using dimensional
analysis, the reader   can easily
restore the hidden factors of $\kappa$ in our expressions.
 In some cases, when $\kappa$ 
is not accompanied by $z$,  we   restore  $\kappa$ explicitly.)
The eigenvalue equation for wave functions $ \psi_n(z) $ of the normalizable modes can be obtained from
Eq.~(\ref{genbasiceq}) by requiring $ p^2 = M_n^2 $, which gives
\begin{align}\label{eom}
\partial_z\left[\frac{1}{z}\,e^{-z^2} \partial_z\, \psi_n \right] + M^2_n\,
\frac{1}{z}\, e^{-z^2}\psi_n = 0 \ .
\end{align}
As noted in Ref.~\cite{Karch:2006pv},    the substitution
\begin{align}\label{sabst}
\psi_n(z) = e^{z^2/2}\sqrt{z}\, \Psi_n(z)
\end{align}
gives a   Schr\"{o}dinger  equation
\begin{align}
\label{schr}
-\Psi''_n + \left(z^2 + \frac{3}{4z^2}\right)\Psi_n = M^2_n \, \Psi_n \
,
\end{align}
which happens to be exactly solvable. The    resulting   spectrum is
 $ M^2_n = 4(n+1) $ (with $ n = 0,1, \dots $),  and
the  solutions $\psi_n(z)$ of the original equation (\ref{eom})
are  given   by
\begin{align}
\label{vnz}
\psi_n(z) = z^2\sqrt{\frac{2}{n+1}} \, L^1_{n}(z^2) \ ,
\end{align}
where $L^1_{n}(z^2)$   are Laguerre  polynomials.
The functions $\psi_n(z)$
are normalized according   to
  \begin{eqnarray}\label{orthov}
   \int_0^\infty \frac{dz}{z} \, e^{-z^2} \, \psi_m(z)\, \psi_n (z) = \delta_{mn}  \ .
  \end{eqnarray}
Correspondingly,  the  $\Psi_n (z) $   functions of the  Schr{\"o}dinger   equation (\ref{schr})
are   normalized  by
 \begin{eqnarray}\label{orthoPsi}
   \int_0^\infty dz  \,  \Psi_m(z)\, \Psi_n (z) = \delta_{mn}  \  ,
  \end{eqnarray}
i.e., just  like wave functions of  bound states in  quantum mechanics.
Note,  however,   that the   functions $\Psi_n (z)$  behave like
$z^{3/2}$   for   small   $z$,  while quantum-mechanical
 wave functions of   bound states with  zero   angular
momentum   have   finite  non-zero  values
at the origin.

%======================================================================

\section{Bulk-to-boundary propagator}

It is   convenient   to represent   $ \tilde V^a_\mu (p, z)$
as  the product  of the 4-dimensional boundary   field $ \tilde V^a_\mu (p)$
and the
bulk-to-boundary propagator $ {\cal  V} (p,z) $  which  obeys  the  basic  equation
\begin{align}
\label{basiceq}
\partial_z\left[\frac{1}{z}\, e^{-z^2}\partial_z {\cal V}\right] + p^2
\frac{1}{z}\, e^{-z^2}{\cal  V} = 0
\end{align}
 that follows from  Eq.~(\ref{genbasiceq})  and satisfies
the boundary condition
 \begin{align}
\label{vp0}
{\cal V}(p,z=0)=1 \  .
 \end{align}
 Its  general solution is  given by  the confluent hypergeometric functions of the first and
second kind
\begin{align}
\label{Vsol}
{\cal V}(p,z)  = A \ {_1}F_1(a, 0, z^2) + B \ U(a, 0, z^2) \ ,
\end{align}
where $ a =  -p^2/4 \kappa^2$, $ A $ and $ B $ are constants.
Since the function ${_1}F_1(a, 0, z^2)$  is  singular for $ z = 0$,
 we    take \mbox{$A=0$.}  Then, for   $a>0$,   the   bulk-to-boundary
propagator ${\cal V}(p,z)$
can   be      written as
 \begin{align} \label{Vpz0}
   {\cal V}(p,z)  =  a \int_0^1 dx  \, x^{a-1} \, \exp \left [ - \frac{x}{1-x} \, z^2  \right ]  \  .
  \end{align}
It is  easy   to   check that this expression satisfies Eqs.~(\ref{basiceq}) and  (\ref{vp0}).
Integrating   by   parts  produces the representation
\begin{align} \label{Vpz}
{\cal V}(p,z)  = z^2  \int_0^1 \frac{dx}{(1-x)^2} \,  x^a \,  \exp \left [ - \frac{x}{1-x}\, z^2 \right ]    \ ,
\end{align}
from   which it  follows that
    if  $ p^2 = 0 $  (or $ a = 0 $),  then
\begin{align}
 {\cal V}(0,z)
= 1
\end{align}
for   all $z$.
The integrand of Eq.~(\ref{Vpz})   contains the  generating function
\begin{align}
 \frac{1}{(1-x)^2} \,   \exp \left [ - \frac{x}{1-x}\, z^2 \right ]   =  \sum_{n=0}^{\infty}
L_n^1 (z^2) \, x^n
\end{align}
for the Laguerre polynomials $L_n^1 (z^2) $,  which gives  the   representation
\begin{align}
\label{btb}
{\cal V}(p,z)  = z^2   \sum_{n=0}^{\infty}
\frac{L_n^1 (z^2)}{a+n+1}
\end{align}
that  can be   analytically  continued   into   the timelike \mbox{$a<0$}   region.
One   can see   that  ${\cal V}(p,z)$  has     poles  there at   expected locations 
$p^2= 4(n+1) \kappa^2$.

The same   representation   for  ${\cal V}(p,z)$   can   be obtained from  the Green's function
\begin{align}
G(p;z,z') = \sum^{\infty}_{n = 0}\frac{\psi_n(z)\psi_n(z')}{p^2 - M^2_n} \
\end{align}
corresponding to
Eq.~(\ref{basiceq}),   namely,
\begin{eqnarray} \label{bulk-boundary}
{\cal V}(p,z') &=& - \left[\frac{1}{z}e^{-z^2}\partial_z
G(p;z,z')\right]_{z = \epsilon \rightarrow 0}   \\[3pt]
&=&  -\sum^{\infty}_{n =
0}\frac{\sqrt{8(n+1)}\psi_n(z')}{p^2 - M^2_n} = -4 \sum^{\infty}_{n =
0} \frac{z^{'2} L^1_n(z^{'2})}{p^2 - M^2_n}  \ ,  \nonumber
\end{eqnarray}
which coincides  with Eq.~(\ref{btb}).

The two-point density function can  also  be obtained   from   the Green's   function:
\begin{eqnarray}
\label{sigma}
\Sigma(p^2) &=&
\frac{1}{g^2_5}\left[\frac{1}{z'}e^{-z'^2}\partial_{z'}\left[\frac{1}{z}e^{-z^2}\partial_z
G(p;z,z')\right]\right]_{z,z' = \epsilon \rightarrow 0}  \nonumber \\ &=&
\sum^{\infty}_{n = 0} \frac{f^2_{n}}{p^2 - M^2_n} \   ,
\end{eqnarray}
where the coupling constants   $ f_{n} =  \kappa^2 \sqrt{8(n+1)}/g_5 $  obtained
 in \cite{Karch:2006pv}  are  determined by
 \begin{eqnarray}
\label{fn}
f_n = \left.
\frac{1}{g_5z}\, e^{-z^2} \, \partial_z  \psi_n (z)
\right|_{z = \epsilon \rightarrow 0}  \   .
\end{eqnarray}
 The  propagator ${\cal V}(p,z$) can   be   represented now  as
\begin{eqnarray} \label{bb2}
{\cal V}(p,z) = g_5  \sum^{\infty}_{n =
0} \frac{f_n \,  \psi_n (z) }{ M^2_n  -  p^2}  \ ,
\end{eqnarray}
where
$
 \psi_n (z)
$
are the original  wave   functions  (\ref{vnz})  corresponding   to  the   solutions
of the   eigenvalue  equation (\ref{eom}).

Given    the   structure  of  Eq.~(\ref{fn}),  it is natural to introduce
 the conjugate
wave functions
\begin{eqnarray}\label{wave-function}
\phi_n(z) &\equiv &  \frac{1}{M_n z}e^{-z^2}\partial_z \psi_n(z) \nonumber \\  &=&
\frac{2}{M_n}e^{-z^2}\left[L^1_n(z^2) - z^2 L_{n-1}^2(z^2)\right] \
,
\end{eqnarray}
whose  nonzero  values  at the origin $ f_n g_5/M_n $ are  proportional  to
the  coupling constant $f_n$ (in this particular   case,  $ f_n g_5/M_n = 
 \sqrt{2} \kappa $).
The inverse   relation between the $\psi$
and $\phi$ wave functions
\begin{align}\label{dual-function}
\psi_n(z) = -\frac{1}{M_n}ze^{z^2}\partial_z\phi_n(z)
\end{align}
 can be obtained from Eq.~(\ref{eom}).    The  $\phi$-functions   are   normalized  by
  \begin{eqnarray} \label{orthophi}
   \int_0^\infty   {dz}\,  {z} \, e^{z^2} \, \phi_m(z)\, \phi_n (z) = \delta_{mn}  \  .
  \end{eqnarray}
In   particular, for   the lowest   states, we have
\begin{align} \phi_0(z) =
\sqrt{2}e^{-z^2}   \   \   \   ,  \   \   \   \phi_1(z) =
\sqrt{2}e^{-z^2}(1-z^2)   \  .
\end{align}
Just like  zero angular momentum  oscillator  wave  functions in quantum
mechanics, these   functions have   finite   values  at $z=0$.
They   also have a Gaussian fall-off $e^{-z^2}$ for   large $z$.
To  make a   more close analogy   with   the oscillator
wave functions,  it  makes sense to absorb the weight
$e^{z^2}$ in Eq.~(\ref{orthophi}) into the wave functions,
i.e., to
  introduce   ``$\Phi$''  wave  functions
\begin{align}
\label{Phiwf}
 \Phi_n(z) \equiv
e^{z^2/2}   \phi_n(z)  =  \frac{1}{M_n z}e^{-z^2/2}\partial_z \psi_n(z)  \ ,
\end{align}
which are   nonzero  at $z=0$,  decrease   like $e^{-z^2/2}$ for   large $z$,
and are normalized according to
 \begin{eqnarray} \label{orthoPhi}
   \int_0^\infty   {dz}\,  {z} \,   \Phi_m(z)\, \Phi_n (z) = \delta_{mn}  \  .
  \end{eqnarray}
The presence  of the $z$  weight in this  condition
(which   cannot   be absorbed into wave functions   without
spoiling their behavior at $z=0$)  suggests   that
pursuing the  analogy   with quantum mechanics  one should
treat $z$
 as  the  radial  variable  of a  2-dimensional
quantum mechanical system.

%======================================================================

\section{3-Point Function}

\noindent The  variation of  the  trilinear (in $V$)  term of   the
action
\begin{align}\label{action}
S_{\rm AdS}^{(3)}  &=
 -\frac{\epsilon_{abc}}{2g^2_5}\int d^4 x
 \int^{\infty}_{\epsilon}\frac{dz}{z}\, e^{-z^2}
 \left(\partial_{\mu}V_{\nu}^a\right)V^{\mu,b}V^{\nu,c}
\end{align}
calculated on the    solutions  of the basic equation
(\ref{basiceq})  gives   the   following result  for   the 3-point
correlator:
\begin{eqnarray} \label{cor}
\langle J_a^{\alpha}(p_1)J_b^{\beta}(-p_2) J_c^{\mu}(q)\rangle &=&
\epsilon_{abc} \, (2\pi)^{4} \, \frac{2i}{g^2_5} \, \delta^{(4)}(p_1
- p_2 + q) \nonumber \\ & &  \hspace{-1cm} \times  \, T^{\alpha \beta \mu} (p_1,p_2,q)  \, W(p_1,p_2,q) \  ,
\end{eqnarray}
with    the   dynamical   part   given   by
\begin{align}
\label{Fsuv} W(p_1,p_2,q) \equiv \int^{\infty}_{\epsilon}
{\frac{dz}{z}}\, e^{-z^2} {\cal V}(p_1,z){\cal V}(p_2,z){\cal V}(q,z) \  ,
\end{align}
and  the   kinematical  factor having  the   structure   of  a nonabelian  three-field  vertex:
\begin{align}
 \label{Vabm}
T^{\alpha \beta \mu} (p_1,p_2,q)&=  \eta^{\alpha \mu }(q -
p_1)^{\beta}
 - \eta^{\beta \mu }(p_2 + q)^{\alpha} \nonumber \\
&+ \eta^{\alpha\beta} (p_1 + p_2)^{\mu} \  .
\end{align}
Incorporating the representation Eq.~(\ref{bb2})  for  the  bulk-to-boundary propagators
gives  the expression
\begin{align}
\label{threepoint} & T(p^2_1, p^2_2, Q^2) = \sum_{n,k =
1}^{\infty}\frac{f_{n} f_{k}  F_{nk}(Q^2) }{\left(p_{1}^2 -
M^2_{n}\right)\left(p_{2}^2 - M^2_{k}\right)}
\end{align}
for  $ T(p^2_1, p^2_2, Q^2) \equiv  W(p_1,p_2,q)/g^2_5 $
as a sum   over the  poles  of  the bound states  in  the  initial and final states.
In  the $z$-integral of Eq.~(\ref{Fsuv}), the   contribution of each bound state
 is   accompanied  by  its   wave
function  $\psi_n(z)$,  while   the  $q$-channel is represented by ${\cal J}(Q,z) = {\cal V}(iQ,z)$.
This gives the  $Q^2$-dependent coefficients
\begin{equation}\label{formfac}
F_{nk}(Q^2)  = \int^{\infty}_{0} \frac{dz}{z} \, e^{-z^2} {\cal
J}(Q,z) \, \psi_n(z)  \,  \psi_k(z)   \ ,
\end{equation}
which   have  the   meaning of     transition  form   factors.
Note  that  since $ {\cal J}(0,z) = 1 $,
 the orthonormality   relation (\ref{orthov})   assures that  $F_{nn}(Q^2=0) =1$
for   diagonal  transitions  and $F_{nk}(Q^2=0) =0$   if $n \neq k$.

 The factor  ${\cal J}(Q,z)$ can     be written as a  sum of monopole
contributions    from the infinite tower of vector mesons:
\begin{equation}
\label{Jmeson} {\cal J}(Q,z) =
 g_5  \sum_{m = 1}^{\infty}\frac{ f_{m}
\psi_m( z)}{ Q^2 + M^2_{m} } \ .
\end{equation}
This decomposition,   discussed in  Refs. \cite{Hong:2004sa,Grigoryan:2007vg},
directly   follows from  Eq.~(\ref{bb2}).
As a  result, the   form   factors $F_{nk}(Q^2)$
can   be   written   in    the   form of a
generalized VMD representation:
\begin{equation}\label{GVMD}
{ F}_{nk}(Q^2) =  \sum_{m=1}^{\infty} \frac{  F_{m,nk} }{ 1+ Q^2/
M_m^2}  \ ,
\end{equation}
where   the   coefficients $F_{m,nk}$  are   given   by    the
overlap   integrals
\begin{equation}
\label{fmnk}  F_{m,nk} = \frac{g_5 f_m}{M_m^2}  \int^{\infty}_{0} \frac{dz}{z} \, e^{-z^2}
\psi_m (z)  \, \psi_n(z)  \,  \psi_k(z) \  .
\end{equation}
%

%======================================================================

\section{Form Factors}

In terms of the $\Psi$ wave  functions of the Schr{\"o}dinger equation (\ref{schr}), the form factors are given by
\begin{equation}\label{formfacPsi}
F_{nk}(Q^2)  = \int^{\infty}_{0}{dz} \,  {\cal
J}(Q,z) \, \Psi_n(z)  \,  \Psi_k(z)   \ ,
\end{equation}
which looks  like an  expression for form factors in quantum   mechanics.
However,   as we  discussed above,  the $\Psi$  wave   functions
are not   direct analogues   of quantum   mechanical
wave functions.   For   such  an analogy,   the $\Phi$   wave functions
(\ref{Phiwf})
are much   more  suitable objects.
So,   let  us introduce form   factors involving $\Phi$
wave functions
\begin{align}
 {\cal F}_{nk} (Q^2) \equiv   \int^{\infty}_{0} dz \, z \,
 {\cal J}(Q,z) \, \Phi_n(z)  \Phi_k(z)  \  .
\end{align}
Again, since $ {\cal J}(Q=0,z) =1$ for all $z$,
the normalization   condition (\ref{orthoPhi})
for the $\Phi_n (z)$ wave functions
guarantees  that  the diagonal form   factors
${\cal F}_{nn} (Q^2)$ are  normalized
to 1  for $Q^2=0$, while the non-diagonal
ones  vanish when  $Q^2=0$.
To establish connection with $F_{nk}(Q^2)$  form   factors,
we use Eq.~(\ref{Phiwf})   to substitute $\Phi$ functions  by
derivatives of $\psi$   wave   functions,
which   gives
\begin{equation}\label{MMF}
M_n M_k {\cal F} _{nk}(Q^2)  = \int^{\infty}_{0} \frac{dz}{z} \, e^{-z^2} {\cal
J}(Q,z) \, \psi_n'(z)  \,  \psi_k'(z)   \ .
\end{equation}
Integrating  $\psi_k'$ by   parts, taking into  account that  \mbox{$\psi_k(0)=0$}  and
incorporating
the  eigenvalue  equation (\ref{eom}) for $\psi_n$ gives
\begin{align}
\label{MMF1}
M_n M_k {\cal F} _{nk}(Q^2) &= M_n^2 { F}_{nk}(Q^2) \\[2pt]
& - \int^{\infty}_{0} \frac{dz}{z} \, e^{-z^2} \, \psi_n'(z)  \,
  \psi_k (z) \,  \partial_z {\cal J}(Q,z) \  . \nonumber
\end{align}
Similarly, integrating $\psi_n'$ by   parts we obtain
\begin{align}
\label{MMF2}
M_n M_k {\cal F} _{nk}(Q^2) &= M_k^2 { F}_{nk}(Q^2) \\[2pt]
& - \int^{\infty}_{0} \frac{dz}{z} \, e^{-z^2} \, \psi_n(z)  \,
  \psi_k'  (z) \,  \partial_z {\cal J}(Q,z) \  . \nonumber
\end{align}
Adding  these two  expressions, integrating $(\psi_n \psi_k)'$
by   parts and  using the basic equation
(\ref{basiceq})  for ${\cal J}(Q,z)$ gives
\begin{align}
F_{nk}(Q^2) &= \frac{2 M_n M_k}{ Q^2 + M^2_n + M_k^2} \,
 {\cal F}_{nk}(Q^2)  \ .
\end{align}
For the  case of diagonal $ n \rightarrow n $ transitions this   gives
\begin{align}
F_{nn}(Q^2) &= \frac{{\cal F}_{nn}(Q^2)}{1 + Q^2/2M^2_n} \  ,
\end{align}
expression   similar
to   that derived in Ref.~\cite{Grigoryan:2007vg}.

Thus,   we   can    obtain
$ F_{nk}(Q^2)$  form   factors from
 the   basic form   factors ${\cal F}_{nk}(Q^2)$.
Note, that these form factors  also  have a  generalized
VMD representation
\begin{equation}\label{GVMDcal}
{\cal F}_{nk}(Q^2) =  \sum_{m=1}^{\infty} \frac{ {\cal F}_{m,nk} }{ 1+ Q^2/
M_m^2}  \ ,
\end{equation}
with   the   coefficients ${\cal F}_{m,nk}$    given   by    the
overlap   integrals
\begin{equation}
\label{Fmnk}  {\cal F}_{m,nk} =
\frac{g_5 f_m}{M_m^2}  \int^{\infty}_{0} {dz}\, {z} \,
\psi_m (z)  \, \Phi_n(z)  \,  \Phi_k(z) \  .
\end{equation}

For  the  lowest diagonal transition (i.e., for  $ n = k = 0 $)   we   have
\begin{align}
\label{calf00gen}
{\cal F}_{00} (Q^2)  =
{2} \int^{\infty}_{0} dz \, z \, e^{-z^2} \,
{\cal J}(Q,z)
\ .
\end{align}
Incorporating  the   representation  (\ref{Vpz})  for  ${\cal J}(Q,z)$   and using
 $a=Q^2/4\kappa^2$,
we  obtain
\begin{align}
\label{calf00}
{\cal F}_{00} (Q^2)  = \frac1{1+a} = \frac1{1+Q^2/M_0^2} \  .
\end{align}
Here,  we took into account  that   the mass of
the lowest   bound   state (i.e., $\rho-$meson)  is $ M_0 = M_{\rho} = 2 \kappa$.

Notice,  that
we   obtained exact vector   meson dominance for ${\cal F}_{00} (Q^2)$:
this   form    factor   is   completely determined by  the  lowest bound   state in  the $q$-channel.
The   higher states  do   not   contribute   because  the   overlap
integral   ${\cal F}_{m,00}$  corresponding to   the   contribution
of the  $m^{\rm th}$  $q$-channel  bound   state  vanishes  for $m>0$:
\begin{align}
{\cal F}_{m,00}   =  2  \int_0^\infty dz \, z^3 \, e^{-z^2} \, L_m^1 (z^2) =\delta_{m0} \  .
\end{align}

In   the case  of   $F_{00}(Q^2)$   form   factor,   we  have
\begin{eqnarray}
F_{00} (Q^2) &=&  \frac{1}{(1+a)(1+a/2)}
\nonumber
\\ &=&  \frac2{1+Q^2/M_0^2} -  \frac1{1+Q^2/M_1^2} \  .
\end{eqnarray}
Thus,  the $F_{00}(Q^2)$   form   factor is   given   by   contributions from the
lowest  two  $q$-channel bound  states.   Since
  $ F_{00} (Q^2)
\sim 1/Q^4 $   for large $ Q^2 $,  exact VMD is impossible
for  this form factor: other resonances are needed to
``conspire''  to    cancel their leading $1/Q^2$
terms  at   large $Q^2$.
In the soft-wall model, this    cancellation is  provided by just the first excited state.

For small $ Q^2 $, the  form   factor $F_{00}^{\rm S} (Q^2)$  has
the following  expansion:
\begin{align}
\label{f00}
F_{00} (Q^2) =   \left[1
-\frac{3}{2}\frac{Q^2}{M^2_{0}} + \frac{7}{4}\frac{Q^4}{M^4_{0}} +
{\cal O}(Q^6) \right] \ .
\end{align}

The  Lorentz structure of   the 3-point  function
in   the soft-wall model    is    the   same as in   the hard-wall model 
considered in Ref.~\cite{Grigoryan:2007vg},  where   it   was
shown that
 electric $G_C$,  magnetic
$G_M$ and quadrupole $G_Q$ form factors
 (for   definitions, see, e.g., \cite{Arnold:1979cg,Grigoryan:2007vg})
of the $n^{\rm th }$
bound state are all expressed
through the $F_{nn}(Q^2)$ form   factor:
\begin{align}
&G_Q ^{(n)}(Q^2) = - F_{nn}(Q^2) \ , \ \ G_M^{(n)} (Q^2) = 2
F_{nn}(Q^2)\ , \\[7pt] \nonumber &G_C^{(n)} (Q^2) = \left  (1-
\frac{Q^2}{6M^2} \right )F_{nn}(Q^2) \ .
\end{align}
The same relations hold for the soft-wall model.
As a result,  small-$Q^2$  expansion of    the electric form factor
of the lowest   bound   state  in   the
soft-wall model    is   given   by
\begin{align}
G_{00}(Q^2) &= \left[1 - \frac{Q^2}{6M^2_{0}}\right]F_{00}(Q^2)
\nonumber \\
&=
\left[1 - \frac{5}{3}\frac{Q^2}{M^2_{0}} + 2\frac{Q^4}{M^4_{0}} +
{\cal O}(Q^6)\right] \ ,
\end{align}
and the electric radius for the $ \rho$-meson
in   the  soft-wall model   is
\begin{align}
\label{radius}
  \langle r^2_{\rho} \rangle^{\rm S}  =
0.655 {\rm \ fm}^2  \  .
\end{align}
  This radius is larger than
the   value  \mbox{$ \langle r^2_{\rho} \rangle ^{\rm H}  =
0.53 {\rm \ fm}^2  $}   that   we  obtained in  Ref.~\cite{Grigoryan:2007vg}
in the case of the hard-wall cutoff.

The radius  of the $n^{\rm th}$ excited  state  can be found from the slope
of $F_{nn}(Q^2)$. The latter   can be calculated using Eqs.~(\ref{vnz}), (\ref{formfac}).
Defining the slope coefficient $S_n$ by
\begin{align}
 \left.  \frac{d}{dQ^2} \,  F_{nn}(Q^2) \right |_{Q^2=0}  \equiv - \frac{S_n}{M_0^2}
\end{align}
and using explicit form of Laguerre polynomials, we find 
\begin{align}
\label{sn1}
S_n &=
  \sum_{m,l=0}^n  C_{n+1}^{m+1} C_{n+1}^{l+1}
  {(-1)^{l+m}} \frac{(m+l+1)!}{(n+1) \, m! \, l!}
\sum_{p=1}^{m+l+2} \frac1p 
\end{align}
($C_\alpha^\beta$ are binomial coefficients). 
A faster algorithm for numerical   calculations is provided by the formula
\begin{align}
S_n &=
  \sum_{m=0}^nC_n^m C_{m+n+1}^{n}
 \sum_{k=0}^{n-m}C_{n-m}^k {(-2)^k}
\sum_{p=1}^{2m+k+2} \frac1p \ .
\end{align}
For $n=0$, these  expressions  give  the result $S_0 =3/2$ corresponding to  Eq.~(\ref{f00}).
For higher states, we have
$S_1={23}/{12}$, $S_2={11}/{5}$, $S_3 \approx 2.415$,   $S_{10} \approx 3.245$,
 $S_{20} \approx 3.816$,  \mbox{$S_{50} \approx 4.633$, }  $S_{100} \approx 5.281$,
 $S_{150} \approx 5.667$, $S_{200} \approx 5.943$ .  For $n \geq 2$, these  values   are  well approximated
by a simple empirical formula 
\begin{align}
 S_n \approx  \ln\, (n+1) + \frac23 + \frac5{4(n+1)} \ .
\end{align}

Thus, the  squared  sizes of excited states increase with the excitation number $n$. However,  contrary to
expectations of Ref.~\cite{Karch:2006pv}, the raise is only  logarithmic, 
\mbox{$  \langle r_n^2 \rangle^{\rm S}  \sim
\ln n$}  rather  than linear.  Such an outcome is  not  unnatural since  Eq.~(\ref{sn1}) differs
from the identity
\begin{align}
  \sum_{m,l=0}^n  C_{n+1}^{m+1} C_{n+1}^{l+1}
  {(-1)^{l+m}} \frac{(m+l+1)!}{(n+1) \, m! \, l!} =1
\end{align}
(that follows from the normalization  condition (\ref{orthoPsi}))
 by the sum
\begin{align}
 \left. \sum_{p=1}^{m+l+2} \frac1p \ \right |_{m+l \to \infty} \sim ~ \ln \, (m+l+2)
\end{align}
which  has a logarithmic behavior for large $m+l$, and  for large $n$  it may be 
approximated by $\ln  n $  for the bulk of $m,l$   values.  However, it would be interesting 
to derive a formal proof.

It should be noted,  that in the hard-wall model , the slope of $F_{nn} (Q^2)$ at $Q^2=0$ {\it decreases}
with $n$. For   the   lowest state, the value $S_1^H = 1.192$ was found in Ref.~\cite{Grigoryan:2007vg}.
For higher radial excitations, we have $S_2^H = 0.877$, $S_3^H = 0.833$, $S_{10}^H = 0.806$,
$S_{20}^H = 0.804$, $S_{100}^H = 0.803$, i.e., 
$\langle r^2_n \rangle^H$  tends to a constant value as $n \to \infty$.

%======================================================================

\section{Comparison with hard-wall model }

Note that in   the  hard-wall  model considered in Ref.~\cite{Grigoryan:2007vg}, all the $q$-channel   states
give  nonzero contributions to ${\cal F}_{00} (Q^2)$.   In   fact, it  is    strongly dominated   by     {\it two}
lowest  $q$-channel   states. The   role   of  the first  excitation  in    the    hard-wall model    is   especially
important   for large $Q^2$:  it   gives     asymptotically $2.061\, M_{\rho}^2/Q^2$ while   the   lowest state
contributes only $0.619 \, M_{\rho}^2/Q^2$.

\begin{figure}[h]
%\hspace{4cm}
\mbox{
  {\epsfysize=5cm  \epsffile{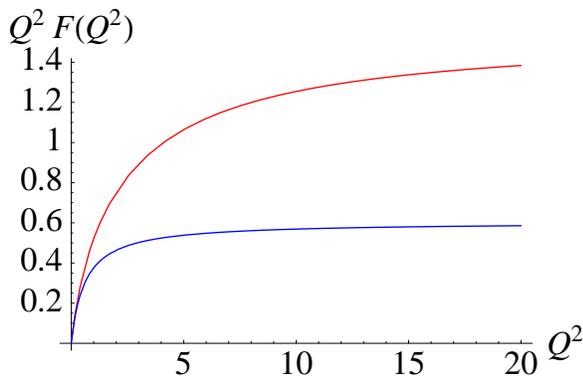}} }
\caption{\label{ff}
$Q^2$-multiplied $\rho$-meson form factor ${\cal F}_{00}  (Q^2)$ (displayed in GeV$^2$) 
as a function of $Q^2$ (given in GeV$^2$) 
in hard-wall (upper line, red online)
and soft-wall (lower line, blue online) models.
}
\end{figure}

It   should   also   be  mentioned that
in   both  models  ${\cal F}_{00} (Q^2)$ has $\sim1/Q^2$   behavior   for   large $Q^2$.
However,
 the   normalization of   the   asymptotic   behavior
in   hard-wall model   is   much larger
than   in     soft-wall model:  ${\cal F}_{00}^{H} (Q^2) \to  2.566 \, M_{\rho}^2/Q^2$,
while ${\cal F}_{00}^{\rm S} (Q^2) \to   \, M_{\rho}^2/Q^2$.

As discussed in Refs.~\cite{Radyushkin:2006iz,Grigoryan:2007vg},
to    calculate the   large-$Q^2$   behavior of ${\cal F}_{00}^{H} (Q^2)$,
one   should take   the large-$Q^2$ limit   of ${\cal J}^{\rm H} (Q,z)$,
which is given by $ zQ K_1 (zQ)\equiv {\cal K} (Qz)  $,
the free-field version of the nonnormalizable   mode.
Asymptotically,  it   behaves  like  $e^{-Qz}$,
so   only  small values of $z$  are important in   the
relevant integral.   As a result,
\begin{equation}
\label{Fas}
 {\cal F}_{00}^{H}  (Q^2) \to \frac{  |\Phi_0^{\rm H}(0)|^2}{Q^2}
\int_0^\infty d\chi \, \chi^2 \,  K_1 (\chi) =  \frac{2 \, |\Phi_0^{\rm H}(0)|^2}{Q^2}  ,
\end{equation}
i.e., the large-$Q^2$   behavior
of ${\cal F}_{00}^{\rm H}  (Q^2)$ is   determined by the   value of the $\Phi$
wave function at  the origin,   which  is   given   by
\begin{equation}
 \Phi_0^{\rm H}(0) = \frac{\sqrt{2}  M_\rho }{ \gamma_{0,1} J_1(\gamma_{0,1})} \approx 1.133 \, M_\rho  \ .
\end{equation}
The nonnormalizable   mode
${\cal J}^{\rm S} (Q,z)$ of  the soft-wall model   should  also   convert
into ${\cal K} (Qz)$ when $Q^2$ is   large.
To   see this  directly, we   compare  the  integral
representation
\begin{align} \label{Kqz}
{\cal K}(Qz)  = z^2  \int_0^1 \frac{dx}{(1-x)^2} \,    \exp \left [ -\frac{(1-x)\, Q^2}{4x}-
\frac{x \,z^2}{1-x}\, \right ]
\end{align}
for  ${\cal K}(Qz) $ and  the representation
\begin{align} \label{Jqz}
{\cal J}^{\rm S}(Q,z)  = z^2  \int_0^1 \frac{dx}{(1-x)^2} \,    \exp \left [- \frac{Q^2}{4}\ln \left (\frac1x \right )
 - \frac{x\, z^2}{1-x} \right ]
\end{align}
for ${\cal J}^{\rm S}(Q,z)$   following from Eq.~(\ref{Vpz}).
For   large $Q^2$,  both    integrals   are dominated by
the region   where \mbox{$1-x \sim 2z\kappa^2/Q$.} Then   both
$(1-x)/x$ and $\ln (1/x)$   may be  approximated by $(1-x)$.
Thus, large-$Q^2$   behavior of ${\cal J}^{\rm S}(Q,z)$
coincides   with that of  ${\cal K}(Qz) $,  and Eq.~(\ref{Fas})  is applicable in soft-wall model  as well,
with   the   normalization of the   asymptotically leading   term
determined by the   value of  $\Phi_0^{\rm S}(z)$   at   the origin,  which is
\begin{equation}
\Phi_0^{\rm S}(0) = M_\rho/\sqrt{2} \approx 0.707\,  M_\rho  \ .
\end{equation}
Hence, it  is  the   difference in   the  values of  $\Phi$  wave   functions
at   the   origin that explains   the difference  in the   asymptotic   normalization
of ${\cal F}_{00}  (Q^2)$ in  these   two   models.

The   difference   in   the values of $\Phi (0)$   leads also  to   difference
in the   values  of   coupling  constants $f_n$   related to $\Phi_n (0)$    by
\begin{equation}
 f_n = \Phi_n (0) M_n /g_5    \  .
   \end{equation}
  The   constant
$g_5$  is   determined  by matching  the  asymptotic
behavior
\begin{align}
\Sigma^{\rm AdS} (p^2) \to - \frac{p^2}{2g_5^2} \, \ln (p^2)
\end{align}
of the  two-point function
$ \Sigma^{\rm AdS} (p^2)$
given   by Eq.~(\ref{sigma})    with the QCD result
for    the  correlator of  the  vector currents $J_\mu = \bar d \gamma_\mu u$
having    quantum   numbers of the $\rho^+$ meson.
Since
\begin{align}
\Sigma^{\rm QCD} (p^2) \to - \frac{N_c}{12\pi^2} \, p^2 \ln (p^2)  \  ,
\end{align}
we   have
\begin{equation}
 g_5 = \sqrt{2} \pi
\end{equation}
for $N_c =3$.   This   gives
\begin{equation}
 f_\rho^{\rm S} = \frac{M_\rho^2}{2 \pi} \approx (309 \, {\rm MeV})^2
\end{equation}
for the $\rho$  coupling   constant  in the soft-wall  model, and
\begin{equation}
 f_\rho^{\rm H} = \frac{M_\rho^2}{ \pi  \gamma_{0,1} J_1(\gamma_{0,1})} \approx (392 \, {\rm MeV})^2
\end{equation}
in  the hard-wall model \footnote{The  hard-wall model    result    $F_\rho^{1/2} \approx 329 \, {\rm MeV}$
 presented in Ref.~\cite{Erlich:2005qh} corresponds to
  the $(\bar  u \gamma_\mu u - \bar  d \gamma_\mu d)/2$  current which   differs by $\sqrt{2}$
from the current $(\bar  u \gamma_\mu u - \bar  d \gamma_\mu d)/\sqrt{2}$ that has the  same normalization   as $
\bar  d \gamma_\mu u$.}. The   experimental value  [22]
\begin{equation}
  f_\rho^{\rm exp}= (401 \pm 4 \, {\rm MeV})^2
\end{equation}
is  very close  to the hard-wall model  result, and in this 
respect the hard-wall model is  more successful.
It may be also noted that,  unlike the  value 
 $ \langle r^2_{\rho} \rangle^{\rm S}  =
0.655 {\rm \ fm}^2$ in Eq.~(\ref{radius}), the  hard-wall model result 
 \mbox{$ \langle r^2_{\rho} \rangle ^{\rm H}  =
0.53 {\rm \ fm}^2  $}  for the $\rho$-meson charge radius obtained in our  paper
\cite{Grigoryan:2007vg} practically coincides  both with  the Dyson-Schwinger model 
result  of Ref.\cite{Bhagwat:2006pu} and lattice gauge calculation
reported in Ref.~\cite{Lasscock:2006nh}.

It is also instructive to  consider the modified coupling
$g_\rho \equiv f_\rho/M_\rho$  that 
has  the dimension of mass, and   determines 
the asymptotical behavior of the form factor. 
 Its value in the soft-wall model 
\begin{equation}
  g_\rho^{\rm S}=  \frac{M_\rho}{2 \pi} \approx 123 \, {\rm MeV}  
\end{equation}
is   close to the experimental value of the pion decay 
constant  $f_\pi \approx 131 \, {\rm MeV}$.  Moreover, 
the pure $\rho$-pole result  (\ref{calf00}) 
is   close to the experimental data on the pion form factor.
So, it is tempting to   take for the pion the same  wave 
functions that were obtained in the $\rho$-meson case
and  use Eq.~(\ref{calf00})   as a model
for the pion form  factor. 
 This was   done 
in  the paper  \cite{Brodsky:2007hb}
(that appeared after we submitted the original
 version  \cite{Grigoryan:2007my}  of the present paper
 to the arxive). 
Taking $\kappa = 375\,$MeV (which is slightly smaller than 
$m_\rho/2$), the authors obtained good agreement 
of the $1/(1+Q^2/4\kappa^2)$ curve with the pion form factor data
(though 
 the value  of $f_\pi^2$ is then  about 30\%  below
 the experimental one).  
However,  within    the model of
 Refs.~\cite{Erlich:2005qh,DaRold:2005zs,Karch:2006pv}, 
which we follow here,
the analysis of the axial-vector current  channel 
requires the inclusion of chiral symmetry breaking effects
absent in the vector current channel.
As a result, wave function equations
for the pion are completely different 
from those for the $\rho$-meson. 
We   discuss the pion form factor in a separate publication
\cite{Grigoryan:2007wn}.

%======================================================================

\section{Summary}

In  the   present  paper,    we  studied      wave functions and   form factors
of vector   mesons   within the framework of   the soft-wall model 
\cite{Karch:2006pv}  which   produces
a more realistic  spectrum  for higher excited mesons \cite{Shifman:2005zn} than   the 
hard-wall model  of  Refs.~\cite{Erlich:2005qh,Erlich:2006hq,DaRold:2005zs}.
Our   analysis   uses  the  approach similar to   that
we  developed in  Ref.~\cite{Grigoryan:2007vg}
in  application to
 the hard-wall model.

An   essential element  of our study of the soft-wall model  is
the   integral representation,  which we   found  for
the  bulk-to-boundary
propagator ${\cal  V}(p,z)$. It    allows
to   write  ${\cal  V}(p,z)$ as an explicit  expansion over
bound state   poles. In   this  sense,  it
plays the same   role as   the Kneser-Sommerfeld
expansion  that we used in our  study \cite{Grigoryan:2007vg} of the hard-wall model.

The pole expansion of ${\cal  V}(p,z)$ involves
 ``$\psi$ wave functions''  that   describe  $z$-dependence of a particular
  pole  contribution
and  are
eigenfunctions of the 5D equation of motion.
However,  since
 $\psi_n(z)$  wave   functions
are   not direct   analogues of the   usual
quantum-mechanical wave   functions,
  we introduced  ``$\Phi$ wave functions''
 resembling  wave
functions of oscillator   states in  quantum
mechanics.    In particular,  the values of these functions
at the origin   give the  couplings $g_5 f_n/M_n$
of   the   bound  states, and   these functions   exponentially
decrease with $z^2$.

Analyzing  the three-point function,
we  obtained   expressions for transition form factors   both
in  terms of the
 $\psi$ wave functions and the ``more physical''
$\Phi$ wave functions.  We  demonstrated  that,  just   like in  the hard-wall model,
the   form factors can   be written in   the   form
of   generalized vector meson dominance
representation, i.e., as   a sum   over all bound
states in   the channel of electromagnetic current
(this  result  confirms the claim  \cite{Hong:2004sa}
that  generalized VMD is a   common feature of AdS/QCD models).

We  derived  an explicit   expression for
 $\rho$-meson form
factors,  and  analyzed  their  behavior  in the   regions of small
and  large $Q^2$.  In particular,  we calculated
the $\rho$-meson electric radius in the soft-wall model,
and   found that it   is larger than in   the hard-wall  model
(the latter  agrees  with calculations in Dyson-Schwinger model \cite{Bhagwat:2006pu} 
and lattice QCD \cite{Lasscock:2006nh}).
Our calculation also demonstrated that the squared
radii of higher excited states increase with $n$,
the number of the radially excited level. However,  contrary to
expectations of Ref.~\cite{Karch:2006pv},
the   increase is only logarithmic rather than linear.
Another result is that, in    the soft-wall model,
the   $\rho$-meson  form   factor
 ${\cal F}_\rho  (Q^2)$  (corresponding to   the   overlap
of   the $\Phi$ wave  functions)
exhibits
an exact  VMD pattern, i.e.,
 it  is given by   a single monopole
term due  to  the  lowest bound state in the
$Q^2$-channel.
In   the case of
the $\rho$-meson form   factor $F_\rho (Q^2)$ (that is given   by   the   overlap
of   the $\psi$-wave  functions),  we found a
 two-resonance dominance  pattern,  when  just  two  lowest bound states  in the
$Q^2$-channel  contribute.

Analyzing the   large-$Q^2$   behavior of the ${\cal F}_\rho  (Q^2)$
form   factor (given by exact $\rho$-pole VMD), we established  that its   asymptotic
normalization in the soft-wall model  is much lower (by  factor 2.566) than
that  of  the hard-wall model.  This difference is  explained by
essentially  lower   value of   the soft-wall model  $\Phi$ wave   function
at the origin.

Finally,   we   calculated the $\rho$-meson   coupling
constant $f_\rho$   both in the soft-wall and hard-wall models, and
found that the  experimental value is  closer  to the hard-wall model  result.

%======================================================================
\vspace{-7mm}

\acknowledgements

\vspace{-3mm} 

H.G. would like to thank J.~Erlich for discussions, A.~W.~Thomas for support at Jefferson Laboratory and
J.~P.~Draayer for support at Louisiana State University. A.R.  thanks  S.~J.~Brodsky and J.~P.~Vary for useful
discussions.

Notice: Authored by Jefferson Science Associates, LLC under U.S. DOE Contract No. DE-AC05-06OR23177. The U.S.
Government retains a non-exclusive, paid-up, irrevocable, world-wide license to publish or reproduce this
manuscript for U.S. Government purposes.

%======================================================================
%======================================================================

\end{document}